\documentclass[%
 reprint, amsmath,amssymb,
 aps,prl]{revtex4-1}

\usepackage{graphicx}
\usepackage{dcolumn}
\usepackage{bm}
\usepackage[colorlinks=true,linkcolor=blue]{hyperref}%

\begin{document}

\title{Fate of electron beam in graphene: Coulomb relaxation or plasma instability?}

\author{Dmitry Svintsov}
 \email{svintcov.da@mipt.ru}
\affiliation{Laboratory of 2d Materials for Optoelectronics, Moscow Institute of Physics and Technology
}

\date{\today}

\begin{abstract}
Electron beams in two-dimensional systems can provide a useful tool to study energy-momentum relaxation of electrons and to generate microwave radiation stemming from plasma-beam instabilities. Naturally, these two applications cannot coexist: if beam electrons do relax, the beam is stabilized; if instability exists, it strongly distorts the distribution function of beam electrons. In this paper, we study the competition of beam relaxation due to electron-electron (e-e) collisions and development of plasma beam instability in graphene. We find that unstable plasma mode associated with a beam is stabilized already by weak e-e collisions. At intermediate e-e collision frequency, the instability re-appears at the ordinary graphene plasmon mode. Such instability is interpreted as viscous transfer of momentum from beam to 2d plasmons. Its growth rate reaches its maximum at hydrodynamic-to-ballistic crossover, when plasmon wavelength and electron mean free path are of the same order of magnitude.
\end{abstract}

\maketitle

The cornerstone of Landau Fermi liquid theory is weak scattering of a single electron excitation over the Fermi surface. The corresponding scattering rate  due to electron-electron (e-e) collisions $\gamma_{ee}$ is proportional to excitation energy squared $\delta\varepsilon^2$ in three dimensions~\cite{Abrikosov_TFL}. In reduced dimensions, the e-e scattering becomes stronger which leads to non-Fermi-liquid behaviour in 1D~\cite{1d_liquids} and log-enhanced scattering ($\sim \delta\varepsilon^2 \ln |\delta\varepsilon|/T$) in 2D~\cite{chaplik_scattering}. Scattering among 2D electrons has recently regained great attention~\cite{Levitov_Collisions} as it leads to novel fluid-like transport observed in numerous experiments~\cite{Levin_NonlocalGaAs, Bandurin_superballistic, Bandurin_Negative, WP2_Hydro}.

While a single electron above Fermi surface inevitably ends up with scattering, the fate of an electron bunch can be far more interesting. Namely, the appearance of a beam over the steady electron background results in a pair of new excitation modes with complex conjugate frequencies, one of which is always unstable (growing)~\cite{Beam_instability_Akhiezer,Physical_Kinetics,Beam_instability}. This phenomenon of plasma beam instability has been actively studied since 1950's in connection with nuclear fission problem~\cite{Parail_tokamak} and solar bursts~\cite{Ginzburg_solar_bursts}.

The resurrection of interest to the fate of electron beams in 2D systems is dictated by two reasons. First, relaxation of injected beam can provide valuable information on the rate of e-e scattering and its energy dependence that is challenging to access with other techniques~\cite{Mollenkamp_beam,Yanovsky_angle-resolved_beam,Jura_beam,Molenkamp_Collimation}. Second, if the plasma-beam instability is more likely than relaxation, such instabilities can form the basis of solid-state terahertz sources~\cite{Bakshi_THz_emission}. The competition between beam scattering and beam instabilities in 2d systems has not been addressed so far, though it was actively studied in three-dimensional Maxwellian plasmas (see review~\cite{Rukhadze_review}).

In this paper, we establish the criteria of beam instabilities and beam relaxation in graphene in the presence of e-e collisions of arbitrary strength. The interest to graphene is motivated by the dominant role of e-e scattering in graphene-based hetersostructures~\cite{Bandurin_Negative,Adam_HD_Window} due to low impurity density and high energy of optical phonons. In addition, formation of high-density electron beams with collimated velocities is easily achievable with graphene tunnel junctions~\cite{Cheianov,Veselago-lens-experiment} and geometrically-patterned contacts~\cite{Bhandari_beams}.

We find that injection of electron beam in graphene leads to the emergence of two plasmon modes one of which is always unstable {\it in the absence of e-e collisions}. Unfortunately and expectantly, already weak e-e scattering in the ballistic regime suppresses the instability. Unexpectedly, we find a new type of instability at the hydrodynamic-to-ballistic crossover where Knudsen number ${\rm Kn} = q v_0 \tau_{ee}$ is order of unity ($q$ is the plasmon wave vector, $v_0$ is the Fermi velocity in graphene, and $\tau_{ee} = \gamma_{ee}^{-1}$ is mean time between e-e collisions). In this regime, the normal graphene plasmons become unstable in the presence of beam, while the physics of instability can be attributed to viscous momentum transfer between beam electrons and collective modes. The effect of e-e collisions on collective modes in this regime is highly non-perturbative, still it can be handled analytically using model collisions integrals~\cite{Abrikosov_TFL,BGK_model,Crossover}.

Before proceeding to calculation, we note a previous attempt to solve the problem of beam instability in graphene in Ref.~\cite{Aryal-instab} in the collisionless case. Even in that case, the results of \cite{Aryal-instab} cannot be called satisfactory as the model ignored Landau damping in graphene at frequencies $\omega < q v_0$. This presence of such damping changes the character of beam instability from threshold-type to thresholdless, as it does in textbook case of Maxwell plasma~\cite{Physical_Kinetics}. Contrary to the case of Maxwell plasmas, there's no small parameter for Landau damping in graphene at frequencies $\omega < q v_0$, and it can by no means be neglected.

\section{Theory of electron beam stability in graphene}

In a canonical problem of plasma-beam instability, the electron beam in collimated in momentum space rather than in real space. The momentum collimation is readily achieved upon electron injection through tunnel junctions~\cite{Cheianov}. The angular distribution of tunnel-injected electrons is Gaussian, and its width shrinks with reducing the barrier transparency. In the following calculations, we shall mimic the distribution function of beam electrons as a delta-function, $f_b({\bf k}) = n_b \delta({\bf k}-{\bf k}_b)$, where $n_b$ is the density of injected electrons with momentum around ${\bf k}_b$, and the delta function is normalized according to $g(2\pi)^{-2}\int{d^2{\bf k} \delta({\bf k}-{\bf k}_0)} = 1$ ($g=4$ is the spin-valley degeneracy). The steady-state distribution function of electrons thus reads 
\begin{equation}
\label{Distribution}
    f_0 ({\bf k})= f_F({\bf k}) + N_b \delta({\bf k}-{\bf k}_0), 
\end{equation}
where $f_F({\bf k})$ is the Fermi function of background equilibrium electrons. 

We are to analyze the electromagnetic stability of distributions \ref{Distribution}. Such analysis is based on evaluation of polarizability $\Pi({\bf q},\omega)$ and dielectric function $\varepsilon({\bf q},\omega)$ of electron system followed by the search of unstable roots for plasmon dispersion relation $\varepsilon({\bf q},\omega)=0$.

The evolution of electron distribution function $f$ is governed by the kinetic equation:
\begin{equation}
 \frac{\partial f}{\partial t} + {\bf v_k}\frac{\partial f}{\partial {\bf r}} + \frac{\partial V}{\partial {\bf r}}\frac{\partial f}{\partial {\bf k}} = \mathcal{C}_{ee}\{f\}
\end{equation}
where ${\bf v_k} = v_0 {\bf k}/k$ is the electron velocity in graphene, $V({\bf r})$ is the external electric potential. The right-hand side is the electron-electron (e-e) collision integral. 

To preserve the main features of e-e collisions and maintain analytical tractability, we adopt $\mathcal{C}_{ee}$ in the generalized relaxation-time approximation. In this model, all perturbations of distribution function are relaxed toward local equilibrium
\begin{gather}
\label{Collision-integral}
\mathcal{C}_{ee}\{f\} = \frac{f - f_{eq}}{\tau_{ee}},\\
f_{eq}({\bf k}) = \left[1+\exp\left\{\frac{\varepsilon_{\bf k} - {\bf ku}_{eq} - \mu_{eq}}{T_{eq}}\right\}\right]^{-1}
\end{gather}
rather than to zero. Moreover, the parameters of this local equilibrium, which are quasi-Fermi level $\mu_{eq}$, drift velocity ${\bf u}_{eq}$ and temperature $T_{eq}$ are different from those of steady background electrons. These parameters would be established after equilibration of background electron plasma and beam, and are determined from particle number, momentum, and energy conservation laws. If the density of beam electrons is small, the equilibrium drift velocity would be $u_{eq} \approx v_0 (n_b/n_0)(k_b/k_F) $.

We further proceed to linearization of Boltzmann equation with respect to small external potential $V({\bf r}) = \delta\varphi_{{\bf q} \omega}e^{i({\bf qr}-\omega t)}$. The distribution function acquires a correction $\delta f_{{\bf q}\omega}({\bf p}) e^{i{\bf qr} - i\omega t}$, and so does the local equilibrium function $f_{eq} = f^{(0)}_{eq} + \delta f_{eq} e^{i{\bf qr} - i\omega t}$. It is now possible to obtain a formal solution for $\delta f_{{\bf q}\omega}({\bf p})$ (the subscript ${\bf q}\omega$ will be suppressed from now on):
\begin{equation}
\label{Delta_f}
\delta f ({\bf k}) = \frac{- {\bf q} e\delta\varphi \frac{\partial}{\partial {\bf k}} \left\{f_F({\bf k}) + f_b({\bf k}) \right\} + i \gamma_{ee} \delta f_{eq}} {\omega + i \gamma_{ee} - {\bf q v_{k}}}.
\end{equation}

Considerable precautions should be taken upon evaluation of momentum derivative for beam distribution function $\partial f_b({\bf k})/\partial{\bf k}$. Once the beam distribution is delta-peaked in momentum space, the derivative becomes ill-defined. This problem is resolved if one recalls that Boltzmann kinetic equation is derived from quantum Liouville equation in the quasi-classical limit. Switching to quantum equations (Appendix A), one finds the replacement rules for pathological terms:
\begin{equation}
\frac{{\bf q} \partial f_b({\bf k})/\partial{\bf k}}{\omega + i \gamma_{ee} - {\bf q v_k}} \rightarrow \frac{f_{b}({\bf k}+{\bf q}) - f_{b}({\bf k})}{\omega + i \gamma_{ee} - \varepsilon_{\bf k + q} + \varepsilon_{\bf k}}.
\end{equation}

The solution for distribution function is accomplished after one finds the parameters of local-equilibrium function
\begin{equation}
\label{Eq-function}
\delta f_{eq} = \delta\mu \partial_\mu f^{(0)}_{eq} + \delta{\bf u} \partial_{\bf u} f^{(0)}_{eq} +\delta T \partial_T f^{(0)}_{eq}.
\end{equation}
using the conservation laws upon collisions. More precisely, the time derivatives of particle number, momentum, and energy should turn to zero if collision integral (\ref{Collision-integral}) is evaluated on distribution functions (\ref{Delta_f}) and (\ref{Eq-function}). This procedure leads us to closed-form equations for local-equilibrium parameters $\delta\mu$, $\delta{\bf u}$, and $\delta T$. These can be called generalized hydrodynamic equations and are valid at arbitrary value of Knudsen number ${\rm Kn} = qv_0 \tau_{ee}$. The final form of these equations is quite cumbersome and presented in Appendix B, yet they yield simple results in hydrodynamic (${\rm Kn} \ll 1$) and ballistic (${\rm Kn} \gg 1$) limits.

\section{Results}
\subsection{Beam instability in graphene: collisionless case}
The polarization $\Pi({\bf q},\omega)$ of electron system with injected beam in the absence of collisions the sum of individual contributions from steady electrons $\Pi_0({\bf q},\omega)$ and the beam $\Pi_b({\bf q},\omega)$. The dielectric function $\varepsilon({\bf q},\omega)$ governing the collective response is therefore
\begin{equation}
\varepsilon = 1 +V_0\left[\Pi_0+\Pi_b\right] \equiv \varepsilon_0 + V_0 \Pi_b,
\end{equation}
where we have introduced the Fourier transform of Coulomb potential in 2D $V_0 = 2\pi e^2/\kappa|q|$ ($\kappa$ is the beackground dielectric constant)~\footnote{One can easily account for the presence of metal gate gate at distance $d$ from graphene by multiplying $V_0$ by $1-e^{-2|q|d}$} and dielectric function of equilibrium graphene electrons $\varepsilon_0 = 1 + V_0 \Pi_0$. The beam polarization is given by
\begin{equation}
\Pi_b = n_b \left[\frac{1}{\omega + i\delta - \omega^{-}_{b0}} - \frac{1}{\omega + i\delta - \omega^{+}_{b0}}\right].
\end{equation}

\begin{figure}[ht]
	\includegraphics[width=0.85\linewidth]{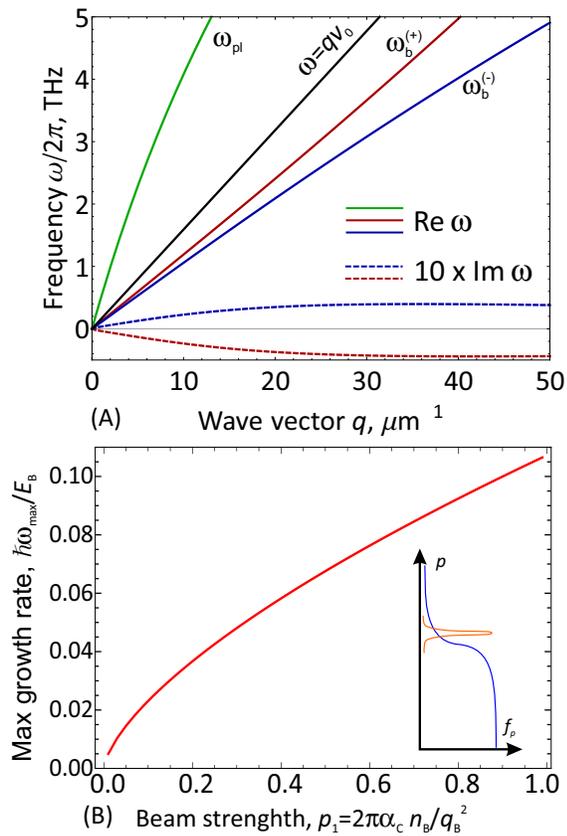}
	\caption{\label{Fig1} 
	Beam instability in collisionless electron system in graphene. Panel (A) shows the calculated dispersion of normal plasmon ($\omega_{\rm pl}$, green line) and two beam-induced modes ($\omega_b^{\pm}$, red and blue lines). Dashed lines show the ten-fold magnified damping and growth rates of beam-induced modes. Panel (B) shows the maximum growth rate of beam-induced mode (scaled by energy of beam electrons) with respect to wave vector $q$, propagation angle $\theta$ and gate-to-channel separation $d$. Inset shows the momentum distribution of electrons in the problem of beam instability
	}
\end{figure}

The polarzation of beam electrons $\Pi_b$ is proportional to their small density, yet it is highly resonant in frequency. The poles of beam polarization are located at two beam-induced collective modes
\begin{equation}
\omega^{\pm}_{b0} = {\bf q v}_b \pm q v_0 \sin\theta \frac{q}{2k_b}.
\end{equation}
The frequency of these modes is {\it almost zero} in the reference frame of a beam, except for a small correction due to quantum effects. Interaction of beam with background electrons through self-consistent field results in modification of beam modes. Their frequencies are changed according to
\begin{equation}
	\label{Beam_modes}
	\omega^{\pm}_{b} = {\bf qv}_b \pm  qv_0|\sin\theta| \left[ \left(\frac{q \sin\theta}{2 k_b}\right)^2 + \frac{V_0 (q) n_b/ k_b}{\varepsilon_0({\bf q},{\bf qv}_b)} \right]^{1/2}.
\end{equation}

The frequencies of these beam-induced modes are no more stable, independent of beam density $n_b$. Indeed, the dielectric function of graphene has non-zero imaginary part at $\omega = {\bf qv}_b < qv_0$ which signifies on collisionless intraband absorption (Landau damping). The double sign before the square root in (\ref{Beam_modes}) implies that one mode ($\omega_b^{+}$) is decaying while the other one ($\omega_b^{-}$) is growing in time.

The beam-induced modes always lie in the domain of intraband absorption $\omega < q v_0$, where the dielectric function of background electrons has positive imaginary part, ${\rm Im}\varepsilon_0 > 0$ (as shown in Fig.~\ref{Fig1}). It may look counter-intuitive that absorptive dielectric function may give rise to plasmon gain, as it follows from Eq.~(\ref{Beam_modes}). This is explained by the fact that the energy of electromagnetic oscillations $W({\bf q},\omega)$ with frequency $\omega_b^{-}$ is negative~\cite{Mikhailovsky}. The time derivative of oscillation energy is negative in absorptive media $dW({\bf q},\omega)/dt<0$, which corresponds to the growth of absolute value of energy.

The plasmon gain appears in the thresholdless manner in the absence of e-e collisions, i.e. even a very small density of beam electrons gives rise to a proportionally small growth rate. The thresholdless character of beam instability is not intrinsic to graphene; already in collisionless warm three-dimensional Maxwell plasma the Landau damping similarly gave rise to a beam instability~\cite{Mikhailovsky}. A special property of graphene (and degenerate 2d electron systems in general) is that Landau damping is never parametrically small, and cannot be neglected in the problem of beam instability. On the contrary, reduction of temperature in Maxwell plasma led to an exponential decrease in Landau damping. In this context, we note that neglect of spatial dispersion in the background dielectric function in the first study of beam instability in graphene was not justified and led to quantitatively wrong conclusions~\cite{Aryal-instab}.   

\begin{figure}[ht]
	\includegraphics[width=0.85\linewidth]{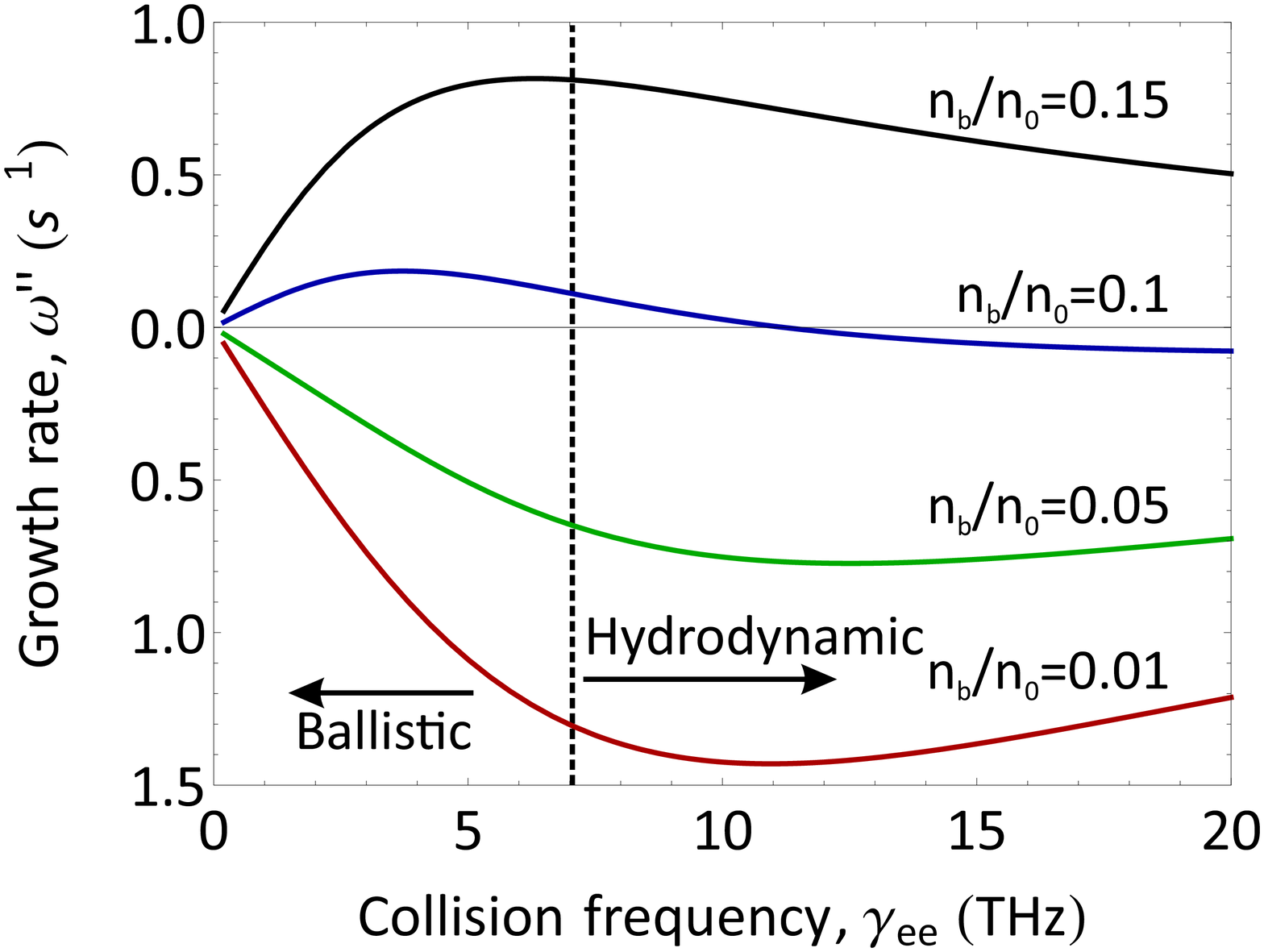}
	\caption{\label{Fig2} 
Excitation of normal graphene plasmons by electron beam at the hydrodynamic-to-ballistic crossover. The figure shows calculated damping/growth rate of normal graphene plasmons vs e-e collision frequency at fixed wave vector  $q = T/(\hbar v_0)$ 	and various densities of beam electrons. For small beam densities, the viscous damping reaches its maximum at ${\rm Kn}\sim 1$, for large beam densities so does the growth rate due to momentum transfer between beam and normal plasmon modes	
	}
\end{figure}

A closer inspection of dispersion relation for unstable modes reveals that the growth rate has maxima as a function of wave vector $q$, propagation angle $\theta$, Fermi energy $\varepsilon_F$ and gate-to-channel separation $d$. The maximum possible growth rate of beam instability can be hardly found analytically, but the outcome of numerical maximization procedure can be presented in universal dimensionless form. In Fig.~\ref{Fig1} B we plot the growth rate in units of $k_b v_0$ vs the dimensionless density of beam electrons $p_1 = 2\pi \alpha_c n_b/k_b^2$. Assuming that beam is injected slightly above the Fermi level, $k_b \approx k_F$, taking the realistically small density of beam electrons $n_b/n_{eq} \approx 0.1$ and coupling constant $\alpha_c =0.5$, we find $p_1 \approx 0.1$ and max growth rate of beam instability $\sim 0.01 \varepsilon_F/\hbar$. It becomes comparable to e-e collision frequency at temperatures $T \approx 0.1 \varepsilon_F$. For realistic Fermi energy $\sim 100$ meV, this corresponds to the liquid nitrogen temperature.

\begin{figure}[ht]
	\includegraphics[width=0.85\linewidth]{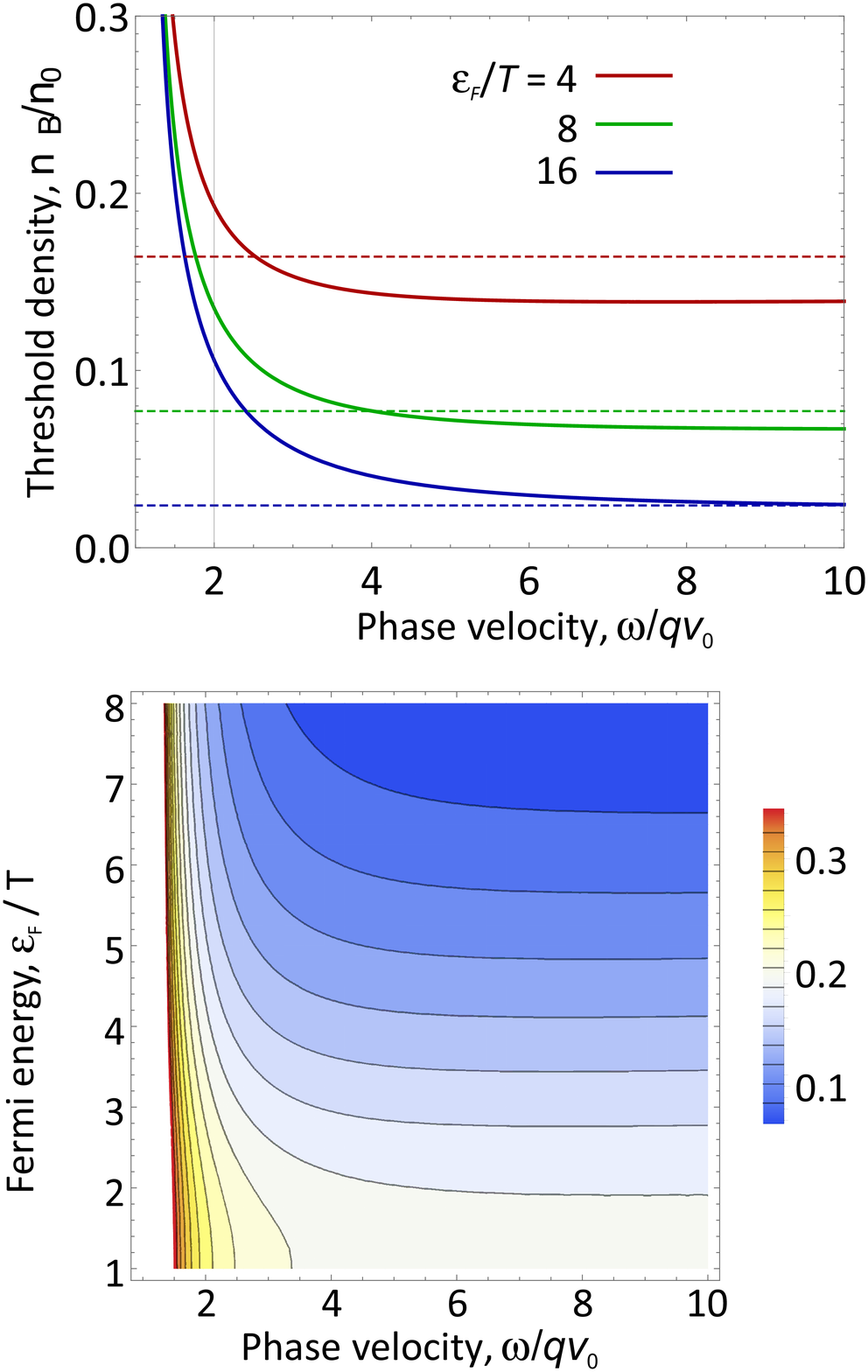}
	\caption{\label{Fig3} 
		Instability of normal graphene plasmons in the nearly-hydrodynamic regime ${\rm Kn} \ll 1$. Panel (A) shows the calculated threshold density of beam electrons for onset of instability vs wave velocity at various Fermi energy (solid lines). Dashed lines show an analytical approximation (\ref{Beam-threshold-limit}) to the threshold density. Panel (B) shows the color map of threshold density vs phase velocity $s$ and Fermi energy $\varepsilon_F/T$
		}
\end{figure}

As the temperature is increased, e-e collisions destroy the ordinary beam instability. As far as the beam density is small ($n_b/n_0 \ll 1$) and collisions can be treated perturbatively ($\gamma_{ee} \ll \omega^{(\pm)}_b$), the effect of collisions is trivial and results in shift of beam mode frequency by $-\gamma_{ee}$. The in-scattering terms of collision integral can be neglected in this regime as they are proportional to the product of beam density and scattering rate. This result can be physically interpreted by smallness of phase space occupied by beam electrons, which results in low probability of electron scattering in the direction of beam propagation. Such conclusion holds for arbitrary model of e-e scattering and is not limited to generalized relaxation time approximation analyzed here.

\subsection{Excitation of graphene plasmons by injected electrons: strong e-e collisions}
The situation changes radically for e-e collisions with frequency comparable to that of plasmon modes. Numerically, it corresponds to terahertz frequencies in graphene at room temperature~\cite{Our-hydrodynamic}. For ordinary plasmon modes in equilibrium with $\omega_{pl} \approx v_0 \sqrt{4\alpha_c k_F q}$, this frequency range is characterized by strong viscous damping~\cite{Crossover}. This result is illustrated in Fig. 2 with the black line. It shows that the damping rate of bulk graphene plasmon vs e-e collision frequency has a maximum located at the crossover between ballistic and hydrodynamic regimes.

When the beam is injected into a electron plasma with strong e-e collisions, the maximum in the damping rate is transformed into the maximum of the growth rate, as shown in Fig. 2 with green and red lines. We have verified that both plasmon damping and beam-induced instability disappear both in the deep hydrodynamic regime ($\gamma_{ee} \rightarrow \infty$) and in the ballistic regime ($\gamma_{ee} \rightarrow 0$). Both effects appear as first-order corrections to plasmon dispersion in Knudsen number ${\rm Kn} = q v_0 \tau_{ee}$; in the absence of the beam the viscous damping equals $\omega'' = qv_0 {\rm Kn}/4$. The fact that beam-induced instability appears at the same order as viscosity enables us to interpret it viscous momentum transfer between electron beam and normal plasmon modes.

The growth rate of 'normal' graphene plasmons due to viscous interaction with electron beam can be studied analytically by expansion of generalized hydrodynamic equations in the limit of small Knudsen number. Noting that the real part of 'normal' plasmon frequency is almost unaffected by scattering, we can obtain the damping/growth rate as:

\begin{equation}
\label{Damping-growth}
\gamma = qv_0 {\rm Kn} \frac{\frac{m_{hd}}{m_{b}} P_1(s,\beta_{eq}) + P_2(s,\beta_{eq}) + n_b P_3(s,\beta_{eq})}{P_4(s,\beta_{eq}) + n_b P_5(s,\beta_{eq})},
\end{equation}  
where $s = \omega/qv_0$ is the phase velocity scaled by Fermi velocity, $\beta_{eq} = u_{eq}/v_0$ is the dimensionless velocity of electrons equilibrated with beam, $m_{hd} = \rho_{eq}v_0^2/n_{eq}$ and $m_{b} = 2 n_{eq}/\partial n_{eq}/\partial\mu$ are the 'proper' electron masses in hydrodynamic and ballistic regimes, and and $P_{i}(s,\beta_{eq})$ are polynomial functions. To the leading order in beam density (and hence, drift velocity) they are given by
\begin{gather}
P_{1} = 4 s \left(1-2
s^2\right)^2 - 4\beta_{\text{eq}} \left(12 s^4+8 s^2-1\right) ,\\
P_{2} = -4 \left(4 s^5-5
s^3+s\right) + 6 \left(6 s^4-8 s^2+1\right) \beta_{\text{eq}},\\
P_{3} = -16 s^5+20 s^3+6 s^2-4 s-2,\\
P_{4} = -32 s^3 + 4\beta_{\text{eq}} \left(26 s^2-1\right), \\
P_{5} = -8 s^3-6 s^2+1.
\end{gather}

In the absence of electron beam, equation (\ref{Damping-growth}) readily reproduces damping rate of equilibrium graphene plasmons~\cite{Crossover}
\begin{eqnarray}
\gamma_0 = - \frac{qv_0 {\rm Kn}}{8}\left[1 + \left(\frac{m_{hd}}{m_{b}} - 1\right)\left(\frac{1}{s}-2s\right)^2\right].
\end{eqnarray}
The first term in square brackets is due to viscous damping. The second one can be traced down to the instrisic conductivity of Dirac fluid~\cite{Gallagher_High-frequency,Quantum_critical,Lucas_resonances,Polini_intrinsic} which, in turn, is a result of non-conserved electric current upon collisions of Dirac electrons.

The beam effects on plasmon growth are proportional to the product of two small quantities, Knudsen number and relative density of beam electrons. It may be questioned whether beam can make a pronounced effect on damping compared to viscosity, which contribution is proportional to $\rm Kn$ solely. It appears that such situation is possible in the limit of large wave phase velocity, $s \gg 1$. In this limit, the viscous damping disappears but the beam-induced growth persists, while the expression for the damping/growth rate acquires a simple form:
\begin{equation}
\gamma_{s\gg 1} \approx - \frac{1}{2} qv_0 s^2 {\rm Kn} \left[\frac{m_{hd}}{m_b} - 1 - n_b\right].
\end{equation}
We observe therefore that beam-induced growth should compete only with damping due to intrinsic conductivity and not with the viscous damping. Further, in the limit of degenerate carriers, $\varepsilon_F/T \gg 1$, the last type of damping disappears, and the threshold density of beam electrons for onset of instability can be relatively small:
\begin{equation}
\label{Beam-threshold-limit}
\frac{N_b}{n_{eq}}\approx \frac{\pi^2}{3}\frac{T^2}{\varepsilon_F^2}.
\end{equation}
The threshold beam density for the onset of plasma instability in the nearly-hydrodynamic regime is a function of only two parameters: dimensionless phase velocity $s = \omega/qv_0$ and scaled Fermi energy. These universal dependences are shown in Fig. 3 with solid lines, dashed lines correspond to analytical low-temperature limits \ref{Beam-threshold-limit}. For small phase velocities $s \sim 1$, the threshold density becomes unachievably large as the beam-induced momentum transfer cannot compensate for viscous dissipation. At large velocities, the instability threshold abruptly goes to zero.

\section{Discussion and conclusions}
The above discussion was concentrated on stability of a spatially uniform distribution comprising steady electrons and collimated beam with energy slightly above Fermi surface. This picture is simplified, as the beam electrons will undergo scattering and angular spreading upon propagation over steady Fermi sea. Within the model adopted model of collision integral, the angular spreading would occur at length $l \sim v_0 \tau_{ee}$. 

More advanced models of scattering in two dimensions account for different relaxation rates for even and odd harmonics of distribution function, $\tau_{\rm even} \approx \tau_{\rm ee} \approx (T/\varepsilon_F)^2 \tau_{\rm odd}$~\cite{Gurzhi_new_effect}. Within these models, the temporal evolution of beam features two characteristic steps~\cite{Gurzhi_beam_relaxation}: (1) angular spreading of electrons across $\delta\theta \sim (T/\varepsilon_F)^{1/2}$ and formation of hole 'tail' in oppostite direction during time $\tau_{\rm even}$ (2) complete angular equilibration during time $\tau_{\rm odd}$. We may suggest that spatial evolution of injected beam would also feature two characteristic lengths, $l_{\rm even} = v_0 \tau_{\rm even}$ and $l_{\rm odd} = v_0 \tau_{\rm odd}$. Stability study of 'pre-equilibrium' beams with angular width $\delta\theta$ is a subject of foregoing research.

It is instructive to compare the criteria of stability for various distributions of drifting electrons. The above study conjectured that electron beam, a distribution of highest possible anisotropy, is unstable in the absence of collisions without any threshold in beam density. Another limiting case is locally-equilibrium distribution of drifting electrons, which represents a Fermi sphere shifted by ${\bf ku}_{\rm dr}$ in momentum space. Such patterns of drifting electrons can lead to instabilities in double-layer and grating-gated graphene, the velocity threshold being $u_{dr} \gtrsim v_0/\sqrt{2}$~\cite{Emission_by_drifting_electrons}. High threshold velocity is paid off by insensitivity of hydrodynamic distributions to e-e collisions, while electron beams are strongly affected by the latter. The instability due to viscous momentum transfer from beam to normal plasmon modes is an appealing exception from this trade-off. 
 
{\it Acknowledgement.} This work was supported by the grant 18-37-20058 of the Russian Foundation for Basic Research. The author thanks Victor Ryzhii for helpful discussions.

\appendix
\begin{widetext}
\section{Generalized hydrodynamic equations}
We repeat the derivation steps of generalized hydrodynamic equations that enable one to find the parameters of local equilibrium distribution function $\delta\mu$, $\delta{\bf u}$, $\delta T$ at given external potential $\delta\varphi$. The details can be found in \cite{Emission_by_drifting_electrons}. The derivation is based on conservation of particle number, momentum, and energy upon e-e collisions. These conservation laws can be symbolically presented as
\begin{equation}
\label{Cons-laws}
\sum_{\bf k}{{\bf g}_{\bf k} \mathcal{C}_{ee}\{\delta f({\bf k})\}} = 0,
\end{equation} 
where ${\bf g}_{\bf k} = \{1, {\bf k},\varepsilon_{\bf k} \}$ is the vector of conserved quantities. With model integral of e-e collisions (\ref{Collision-integral}), the above requirement simplifies to:
\begin{equation}
\label{Cons-laws-2}
\sum_{\bf k}{{\bf g}_{\bf k} [\delta f({\bf k}) - \delta f_{eq}({\bf k})]} = 0,
\end{equation} 
where $\delta f({\bf k})$ is given by (\ref{Delta_f}) and $\delta f_{eq}({\bf k})$ -- by Eq.~(\ref{Eq-function}). Upon performing integration over momentum space, we are led to a system of generalized hydrodynamic equations.

Formulation of this system in terms of $\delta\mu$, $\delta{\bf u}$, and  $\delta T$ is inconvenient. Instead, we pass to the variations of particle density $\delta n$, relativistic velocity $\delta \beta = \delta u/v_0$, and mass density $\delta\rho$. These variations are bound by equations of state:
\begin{gather}
    \delta n = \frac{\partial n}{\partial \mu}\delta\mu + \frac{3 \beta n}{1-\beta^2}\delta\beta + \frac{2n - \mu \partial n/\partial\mu}{T}\delta T,\\
    \delta \rho = (\rho - \mu n/v_0^2) \frac{3\delta T}{T} + \frac{5\beta \rho}{1-\beta^2}\delta\beta.
\end{gather}
Introducing the vector of unknown quantities ${\bf x} = \{\delta n/n, \delta\beta,\delta\rho/\rho\}$, we can formulate the generalized hydrodynamic equations in a symbolic matrix form
\begin{equation}
\label{HD_system}
{\hat M}{\bf x} = {\bf f}_{pl} + {\bf f}_{b},
\end{equation}
where ${\bf f}_{pl}$ and ${\bf f}_{b}$ can be considered as generalized forces acting on background electron plasma and electron beam. The hydrodynamic matrix has the form
\begin{equation}
\label{HD_matrix}
\hat{M}=\left( \begin{matrix}
1-i{{\tilde{\gamma }}_{ee}}{J_{02}} & -i{{\tilde{\gamma }}_{ee}}{{\partial }_{\beta }}{J_{02}} & 0  \\
0 & 1-\frac{2i}{3}{{{\tilde{\gamma }}}_{ee}}{{\partial }_{\beta}}{J_{13}} & {{\beta }_{eq}}-\frac{2i}{3}{{\tilde{\gamma }}_{ee}}{J_{13}}  \\
0 & {\beta_{eq}}-i{{\tilde{\gamma }}_{ee}}{{\partial }_{\beta }}{J_{03}} & 1-i{{\tilde{\gamma }}_{ee}}{{J}_{03}}  \\
\end{matrix} \right),
\end{equation}
where we have introduced the dimensionless strength of e-e collisions $\tilde\gamma_{ee} = (qv_0 \tau)^{-1} = {\rm Kn}^{-1}$ . The dimensionless quantities $J_{nm}$ depend only on equilibrium velocity $\beta_{eq}$ and ratio $a=(\omega+i\gamma_{ee})/qv_0$:
\begin{equation}
J_{nm}\left( a,\beta  \right)=\frac{{{\left( 1-\beta^2 \right)}^{m-\frac{1}{2}}}}{2\pi }\int\limits_{0}^{2\pi }{\frac{{{\cos }^{n}}\theta d\theta }{{{\left( 1-\beta \cos \theta  \right)}^{m}}\left( a-\cos \theta  \right)}}.
\end{equation}
The force vectors have the form
\begin{gather}
	\label{f_plasma}
	{{\bf f}_{pl}}=-2 \left( \begin{aligned}
		& \frac{J_{10}}{{m_{b}}v_{0}^{2}} \\ 
		& \frac{J_{20}}{{m_{hd}v_{0}^{2}}} \\ 
		& \frac{3J_{10}/2}{{m_{hd}v_{0}^{2}}} \\ 
	\end{aligned} \right) ,\\
	\label{f_beam}
	{{\bf f}_{b}}=\frac{n_b}{n}\left( \begin{aligned}
		& \frac{1}{\omega +i{{\gamma }_{ee}}-\omega _{b}^{-}}-\frac{1}{\omega +i{{\gamma }_{ee}}-\omega _{b}^{+}} \\ 
		& \frac{1}{{m_{hd}}}\left( \frac{{k_b}{v_0}\cos \theta }{\omega +i{{\gamma }_{ee}}-\omega_b^{ -}}-\frac{k_bv_0\cos \theta +q{v_0}}{\omega +i{{\gamma }_{ee}}-\omega _{b}^{ + }} \right) \\ 
		& \frac{1}{2m_{hd}/3}\left( \frac{\varepsilon_{k_b}}{\omega +i{\gamma }_{ee}-\omega_b^-}-\frac{\varepsilon_{k_b + q}}{\omega +i{\gamma }_{ee}-\omega _b^+} \right) \\ 
	\end{aligned} \right).
\end{gather}

Though being quite tedious, the system enables full analytical treatment at arbitrary e-e collision frequency.


\end{widetext}

\bibliography{E-beam-biblio}

\end{document}